\documentclass[onecolumn,floatfix,showpacs,aps,nofootinbib]{revtex4}

\usepackage{amssymb,amsmath}
\usepackage{graphicx,float}
\usepackage{indentfirst}
\newcommand{\beq}{\begin{equation}}
\newcommand{\eeq}{\end{equation}}
\newcommand{\bea}{\begin{eqnarray}}
\newcommand{\eea}{\end{eqnarray}}
\newcommand{\nn}{\nonumber}

\def\({\left(}
\def\){\right)}
\def \pt{\partial}
\def \L{\mathcal{L}}
\def \= {\,\dot{=}\,}

\begin{document}

\title{ Simple prescription for  computing  the nonrelativistic interparticle potential energy related to dual models}
\author{G.B. de Gracia$^1$}
\email{gb9950@ift.unesp.br}
\author{G.P. de Brito$^2$}
\email{gpbrito@cbpf.br}

\affiliation{$^1$Instituto de F\'{i}sica Te\'{o}rica (IFT), Universidade Estadual Paulista (UNESP), \mbox{Rua Dr. Bento Teobaldo Ferraz, 271,
Bloco II, Barra Funda, 01140-070, S\~{a}o Paulo, SP, Brazil}}

\affiliation{$^2$\mbox{Laborat\'{o}rio de F\'{i}sica Experimental (LAFEX), Centro Brasileiro de Pesquisas F\'{i}sicas (CBPF),}\\ Rua Dr. Xavier Siguad, 150, Urca, 22290-180, Rio de Janeiro, RJ, Brazil}


\begin{abstract}
Following a  procedure recently utilized by Accioly {\it et al.} to obtain the $D$-dimensional interparticle potential energy  for electromagnetic models  in the nonrelativistic limit, and relaxing the condition assumed by the authors concerning the conservation of the external current,   the prescription found out by them is generalized  so that   dual models can  also be contemplated.  Specific  models  in which the   interaction is mediated by a spin-0 particle described first by a vector field and then by a higher-derivative vector field, are analyzed. Systems mediated by spin-1 particles described, respectively, by symmetric  rank-2 tensors,   symmetric rank-2 tensors augmented by higher derivatives, and  antisymmetric rank-2 tensors, are considered as well.
   
\end{abstract}
\maketitle
\section{Introduction}

Currently, a great number of models modeling interactions have appeared in the literature. Nonetheless, all these systems are required to possess a well defined behavior  in the nonrelativistic limit, whatever their high energy aspects may be. For instance, models describing  electromagnetic interactions must reproduce the Coulomb potential energy plus a possible correction; while gravitational systems must lead to Newtonian gravity sometimes enlarged by a correction.
 Accordingly, any simple method that  could enable us to find  the nonrelativistic potential energy without  expending much effort, should be welcome. Recently, Accioly \textit{et.al.} developed an straightforward   method for addressing  this question \cite{Accioly1,Accioly2}. 

In Ref. \cite{Accioly1},  a suitable prescription for computing the nonrelativistic interparticle potential energy related to $D$-dimensional electromagnetic models described by vector fields is constructed; whereas  in Ref. \cite{Accioly2}, the interparticle potential energy for $D$-dimensional gravitational systems was analyzed. In both articles, the authors assumed that the external current is conserved. 

In this paper our goal is to extend the prescription built out by  Accioly  {\it et al.} \cite{Accioly1,Accioly2} in order to incorporate the possibility of dual theories. Duality, which is the equivalence between different mathematical description of the same physics, has been extensively considered in physics \cite{Duality}. One of the most interesting features of duality is the possibility of changes in the coupling regimes, for instance, a weak (strong) coupling in one model has an strong (weak) coupling in the corresponding model. Hence, provided that we know the correspondence between the original and the dual model, one can perform perturbative computations in both regimes.

In the context of dual theories, a vector field can be associated with both spin-$0$ ou spin-$1$ particle and, in the first case, the longitudinal part of the an external current contributes to the potential energy. In addition, a rank-2 tensor field has several representations and,  in general,  can be  split  in  symmetric ($H_{\mu\nu}$) and antisymmetric ($A_{\mu\nu}$) parts. The symmetric tensor,  $H_{\mu\nu}$, in turn,  gives rise to four different sectors: spin-2, spin-1, and two spin-0 sectors. As a result, we may expect  different kinds of couplings between the symmetric tensor $H_{\mu\nu}$ and external currents. In fact, the dual representation of spin-1 and spin-0 particles described by a symmetric tensor  admits couplings with tensor currents containing components along the longitudinal direction. 

In the aforementioned cases, the prescription developed by Accioly \textit{et al.} \cite{Accioly1,Accioly2} cannot be applied, since the presence of longitudinal components in the external currents is in conflict with the hypothesis of conserved external currents. Therefore, in this paper we intend to extend the prescription in order to incorporate the possibility of external currents with longitudinal components.



The paper is organized as follows. In Sec. \ref{vector_case} we introduce the generalized prescription for the case of interactions mediated by vector fields. In Sec. \ref{section_tensor} we investigate the issue of interactions mediated by a symmetric rank-2 tensor field. In Sec. \ref{antissimetrico} we adapt the prescription developed in the two last sections  to the situation where the interaction is mediated by an antisymmetric rank-2 tensor field.  To text the efficacy and simplicity of the method we have developed,  explicit examples of the computation of the interparticle potential energy concerning the situations dealt with in Secs. II-IV, are presented. 
Finally, in Sec. \ref{conclusion}, we concluded. Technical details are relegated to the Appendices.

Natural units  are used throughout, and our Minkowski metric is $diag(1, -1, ..., -1)$.

\section{Prescription for computing the non-relativistic interparticle potential energy  mediated by  vector fields \label{vector_case}}

It is well known that the $D$-dimensional static potential energy associated with an external vectorial current $J^\mu$ can be obtained from the  expression \cite{Zee}
\begin{equation}
E_D = \frac{1}{2 \tau} \int \int d^Dx \, d^Dy\, J^\mu(x) D_{\mu\nu}(x-y) J^\nu(y),
\end{equation}
where $D_{\mu\nu}(x-y)$ is the Feynman propagator, $J^\mu(x)$ is an external current and the parameter $\tau$ stands for a time interval.  Consider  then the following decomposition for the  external current $J^\mu(x)$, namely,  $J^\mu(x)= J^\mu_T(x) + J^\mu_L(x)$, where $J^\mu_T$ and $J^\mu_L$ are, respectively, the transverse and longitudinal parts of the aforementioned current; of course, only the transverse  part  of the current must necessarily satisfy the conservation law $\partial_\mu J^\mu_T (x) = 0$. Taking  the above decomposition into account, we find
\begin{eqnarray}\label{energy_1}
E_D &=& \frac{1}{2 \tau} \int \int d^Dx \, d^Dy\, J^\mu_T(x) D_{\mu\nu}(x-y) J^\nu_T(y) + \frac{1}{2 \tau} \int \int d^Dx \, d^Dy\, J^\mu_L(x)  D_{\mu\nu}(x-y) J^\nu_L(y)  + \nn\\ &+&  \frac{1}{2 \tau} \int \int d^Dx \, d^Dy\, J^\mu_T(x) D_{\mu\nu}(x-y) J^\nu_L(y) +\frac{1}{2 \tau} \int \int d^Dx \, d^Dy\, J^\mu_L(x)  D_{\mu\nu}(x-y) J^\nu_T(y).
\end{eqnarray}
Bearing in mind that
\begin{equation}\label{propagator_fourier}
D_{\mu\nu}(x-y) = \frac{1}{(2\pi)^D} \int\,d^Dk \, D_{\mu\nu}(k) e^{ik(x-y)},
\end{equation}
we may rewrite  the  potential  energy as follows
\begin{eqnarray}\label{energy_2}
E_D =\frac{1}{2 \tau} \int\,\frac{d^Dk}{(2\pi)^D}  D_{\mu\nu}(k) \int \int d^Dx \, d^Dy\, \bigg\{ J^\mu_T(x) J^\nu_T(y) + J^\mu_L(x) J^\nu_L(y) + J^\mu_T(x) J^\nu_L(y)+J^\mu_L(x)J^\mu_T(y) \bigg\} e^{ik(x-y)}.
\end{eqnarray}
Since we are only interested in time-independent currents, \textit{i.e.}, $J^{\mu}(x) = J^\mu(\textbf{x})$, we integrate in the variables $x^0$, $y^0$ and $k^0$, which leads to
\begin{eqnarray}\label{energy_3}
E_D \!= \!\!\! \int\!\! \frac{d^{D\!-\!1}\textbf{k}}{(2\pi)^{\!D-\!1}} D_{\mu\nu}(\textbf{k}) \left[\Delta^{\mu\nu}_T(\textbf{k}) \!+\! \Delta^{\mu\nu}_L(\textbf{k}) \!+\! \Omega^{\mu\nu}(\textbf{k}) \right] ,
\end{eqnarray}
where we  have used the definitions 
\begin{equation}\label{delta_T}
\mkern-12mu \Delta^{\mu\nu}_T(\textbf{k}) = \frac{1}{2}\int \int \!d^{D-1}\textbf{x} \, d^{D-1}\textbf{y}\,  J^\mu_T(\textbf{x})J^\nu_T(\textbf{y}) e^{i\textbf{k}\cdot(\textbf{y}-\textbf{x})},
\end{equation}
\begin{equation}\label{delta_L}
\mkern-12mu \Delta^{\mu\nu}_L(\textbf{k}) = \frac{1}{2}\int \int \!d^{D-1}\textbf{x} \, d^{D-1}\textbf{y}\,  J^\mu_L(\textbf{x})J^\nu_L(\textbf{y}) e^{i\textbf{k}\cdot(\textbf{y}-\textbf{x})},
\end{equation}

\begin{eqnarray}\label{delta_TL}
\Omega^{\mu\nu}(\textbf{k}) = \frac{1}{2}\int \int \!d^{D-1}\textbf{x} \, d^{D-1}\textbf{y}\,  J^\mu_T(\textbf{x})J^\nu_L(\textbf{y}) e^{i\textbf{k}\cdot(\textbf{y}-\textbf{x})} +\frac{1}{2}\int \int \!d^{D-1}\textbf{x} \, d^{D-1}\textbf{y}\, J^\mu_L(\textbf{x})J^\nu_T(\textbf{y}) e^{i\textbf{k}\cdot(\textbf{y}-\textbf{x})}.
\end{eqnarray}

 After the addition of a  possible gauge fixing term, the most general free electromagnetic Lagrangian (containing only  quadratic terms) that can be constructed may be cast in the form
\begin{equation}\label{free_lagrangian}
\L_0 = \frac{1}{2} A^{\mu} [ a(\Box) \,\theta_{\mu\nu} + b(\Box) \,\omega_{\mu\nu}] A^{\nu} ,
\end{equation}
where $\theta_{\mu\nu} = \eta_{\mu\nu} - \omega_{\mu\nu}$ and $\omega_{\mu\nu} = \pt_{\mu}\pt_{\nu}/\Box$ stand for the usual vectorial  projection operators, and the coefficients $a(\Box)$ and $b(\Box)$ are polynomial functions of the d'Alembertian operator.  To perform the transition from the usual coordinate space to the momentum representation we use the straightforward correspondence: $\Box \to -k^2$ and $F(\Box) \to F(-k^2) \= \tilde{F}(k)$, for any function $F(\Box)$. The Feynman propagator in  momentum space is then obtained directly from the above Lagrangian, as follows
\begin{equation}\label{propagator_vector}
D_{\mu\nu}(k) = \frac{1}{\tilde{a}(k)}\theta_{\mu\nu} + \frac{1}{\tilde{b}(k)}\omega_{\mu\nu} .
\end{equation}
Using the above expression  and taking into account that $\theta_{\mu\nu}J^{\nu}_L = \omega_{\mu\nu}J^\nu_T = 0$, we promptly obtain
\begin{eqnarray}
D_{\mu\nu}(\textbf{k}) \Delta^{\mu\nu}_{T}(\textbf{k}) = \frac{1}{\tilde{a}(\textbf{k})}\eta_{\mu\nu}\Delta^{\mu\nu}_{T}(\textbf{k}),
\end{eqnarray}
\begin{eqnarray}
D_{\mu\nu}(\textbf{k}) \Delta^{\mu\nu}_{L}(\textbf{k}) = \frac{1}{\tilde{b}(\textbf{k})}\eta_{\mu\nu}\Delta^{\mu\nu}_{L}(\textbf{k}),
\end{eqnarray}
\begin{eqnarray}
D_{\mu\nu}(\textbf{k}) \Omega^{\mu\nu}(\textbf{k}) = 0,
\end{eqnarray}
where $\tilde{a}(\textbf{k}) = \tilde{a}(k)|_{k^0 =0}$, $\tilde{b}(\textbf{k}) = \tilde{b}(k)|_{k^0 =0}$. Substitution of the above equations into (\ref{energy_3}), allows to conclude that
\begin{eqnarray}
E_D \!= \! \eta_{\mu\nu}\! \int \frac{d^{D-1}\textbf{k}}{(2\pi)^{D-1}} \left[\frac{\Delta^{\mu\nu}_T(\textbf{k})}{\tilde{a}(\textbf{k})} + \frac{\Delta^{\mu\nu}_L(\textbf{k})}{\tilde{b}(\textbf{k})} \right].
\end{eqnarray}

Accordingly, there are two contributions for the interparticle potential energy, one coming from interactions mediated by spin-0 particles (longitudinal sector) and the other one associated with interactions mediatet by spin-1 particles (transverse sector). It is important to highlight that for  a  model which propagates only one of the sectors (spin-0 or spin-1),  the contribution associated with the other sector must be discarded.  For clarity's sake, we use the following notation
\begin{equation}\label{potential_vector_TL}
E_D = E^{(T)}_D + E^{(L)}_D
\end{equation}
where
\begin{equation}\label{potential_tranversal}
E^{(T)}_D = \eta_{\mu\nu}\! \int \frac{d^{D-1}\textbf{k}}{(2\pi)^{D-1}} \frac{\Delta^{\mu\nu}_T(\textbf{k})}{\tilde{a}(\textbf{k})}\, \leftrightarrow \textmd{spin-1 contribution},
\end{equation}

\begin{equation}\label{potential_longitudinal}
E^{(L)}_D = \eta_{\mu\nu} \int\frac{d^{D-1}\textbf{k}}{(2\pi)^{D-1}} \frac{\Delta^{\mu\nu}_L(\textbf{k})}{\tilde{b}(\textbf{k})}\, \leftrightarrow \textmd{spin-0 contribution}.
\end{equation}

Now, since we are only interested in the computation of the  potential energy associated with two  point-like  static charges, we specify the transverse and longitudinal currents, respectively, as follows
\begin{equation}\label{vector_current_1}
J^\mu_T(\textbf{x})\!=\! \eta^{\mu 0}\!\left(Q^{(T)}_1 \!\delta^{d}(\textbf{x}-\textbf{a}_1)\! +\! Q^{(T)}_2 \!\delta^{d}(\textbf{x}-\textbf{a}_2)\right), 
\end{equation}

\begin{eqnarray}\label{scalar_current_1}
J^\mu_L(\vec{x}) = \pt^\mu\left(Q^{(L)}_1 \delta^{d}(\textbf{x}-\textbf{a}_1)+ Q^{(L)}_2 \delta^{d}(\textbf{x}-\textbf{a}_2)\right),
\end{eqnarray}
where $d=D-1$ is the number of spatial dimensions.

Two comments are in order here.
\begin{enumerate}
\item 
The longitudinal sector may be interpreted as the gradient of a scalar current

\item  $Q^{(T)}$ and $Q^{(L)}$ do not represent charges of the same nature; as a consequence, there is no reason to expect that $Q^{(T)}$ and $Q^{(L)}$ have the same dimension. In fact, the dimension of the charges $Q^{(T)}$ and $Q^{(L)}$ depends on  the dimension of the field which mediates  the interaction. A simple dimensional analysis shows  that the charges $Q^{(T)}$ and $Q^{(L)}$ have, respectively, mass dimension $1-n$ and $-n$, where $n$ stands for the mass dimension of the vector field.

\end{enumerate}

  Substitution of Eqs. (\ref{vector_current_1}) and (\ref{scalar_current_1}), respectively, into (\ref{delta_T}) and (\ref{delta_L}), leads to  
\begin{equation}
\Delta^{\mu\nu}_T(\textbf{k}) = \eta^{\mu 0} \eta^{\nu 0} Q^{(T)}_1 Q^{(T)}_2 \cos(\textbf{k} \cdot \textbf{r}),
\end{equation}

\begin{equation}
\Delta^{\mu\nu}_L(\textbf{k}) = \delta^{\mu}_i \delta^{\nu}_j Q^{(L)}_1 Q^{(L)}_2 k^i k^j \cos(\textbf{k} \cdot \textbf{r}),
\end{equation}
where, $\textbf{r} = \textbf{a}_2 - \textbf{a}_1$. Using the last two equations in (\ref{potential_tranversal}) and (\ref{potential_longitudinal}), we obtain
\begin{equation}\label{potential_vector_T}
E^{(T)}_D = \frac{Q^{(T)}_1 Q^{(T)}_2}{(2\pi)^{D-1}}\int d^{D-1}\textbf{k} \frac{e^{i\textbf{k}\cdot\textbf{r}}}{\tilde{a}(\textbf{k})}.
\end{equation}

\begin{equation}\label{potential_vector_L}
E^{(L)}_D = -\frac{Q^{(L)}_1 Q^{(L)}_2}{(2\pi)^{D-1}}\int d^{D-1}\textbf{k} \frac{\textbf{k}^2\,e^{i\textbf{k}\cdot\textbf{r}}}{\tilde{b}(\textbf{k})}.
\end{equation}
It is worth noting  that in the preceding equations we have replaced the cosine function by an exponential since the integrals are  invariant under the transformation $\textbf{k}\to-\textbf{k}$. We call attention to the fact if the external current is conserved, (22) coincides with the result found in Ref. 1.

Summing up, we have built out  a simple prescription for computing the non relativistic interparticle potential energy mediated by vector fields. To utilize the alluded method we follow the steps below. 
\begin{enumerate}
\item Write the free Lagrangian as in (\ref{free_lagrangian}).
\item Determine the coefficients $\tilde{a}(\vec{k}) = \tilde{a}(k)|_{k^0=0}$ and $\tilde{b}(\vec{k}) = \tilde{b}(k)|_{k^0=0}$.
\item Verify if there is some conservation law related  with  the external  source $J^\mu$ and, if that is the case, exclude the longitudinal or transverse contribution.
\item Determine what mode contributes to the particle content of the theory and, as a consequence, establish what part of the external current couples with the vector field.
\item Compute  the interparticle potential energy  using Eqs. (\ref{potential_vector_TL}), (\ref{potential_vector_T}) and (\ref{potential_vector_L}).
\end{enumerate}

\section{ Prescription for computing the non-relativistic interparticle potential energy mediated by symmetric rank-2 tensor fields \label{section_tensor}}
Since in Secs. III  and IV we will make use of an approach similar to that utilized in the last section, we shall only present in the mentioned sections the main results of the method. In this way, boring repetitions will be avoided. 
 
The potential energy mediated by a tensor field may be calculated from \cite{Zee}
\begin{equation}
E_D \!=\! \frac{1}{2\tau} \int \!\!\int d^Dx\,d^Dy\, J^{\mu\nu}(x)D_{\mu\nu\alpha\lambda}(x-y)J^{\alpha\lambda}(y).
\end{equation}

We appeal now to the convenient tensor decomposition $J^{\mu\nu} = J^{\mu\nu}_{TT} + J^{\mu\nu}_{TL} + J^{\mu\nu}_{LL}$, where $J^{\mu\nu}_{TT}$ has only transverse components and satisfy the conservation law $\pt_\mu J^{\mu\nu}_{TT} = 0$.  In addition, $J^{\mu\nu}_{TL}$ can be written in terms of a transverse vectorial current $J^{\mu\nu}_{TL} = \pt^\mu J^\nu_T + \pt^\nu J^\mu_T$ while $J^{\mu\nu}_{LL}$ has only longitudinal components, which implies that it can be written in terms of a scalar current $J^{\mu\nu}_{LL} = \pt^\mu \pt^\nu J$. It is remarkable that the above decomposition is  quite general  since it preserves all degrees of freedom of $J^{\mu\nu}$. 
 Therefore,
\begin{eqnarray}\label{potential_tensor}
E_D = \int \frac{d^{D-1}\textbf{k}}{(2\pi)^{D-1}}\, D_{\mu\nu\alpha\lambda}(\textbf{k}) \bigg( \Delta^{\mu\nu\alpha\lambda}_{TT}(\textbf{k}) + \Delta^{\mu\nu\alpha\lambda}_{LL}(\textbf{k}) + \Delta^{\mu\nu\alpha\lambda}_{TL}(\textbf{k}) + \Upsilon^{\mu\nu\alpha\lambda}(\textbf{k}) + \Pi^{\mu\nu\alpha\lambda}(\textbf{k})\bigg), \qquad\quad
\end{eqnarray}
where the following suitable definitions were used
\begin{equation}\label{delta_TT}
\mkern-12mu \Delta^{\mu\nu\alpha\lambda}_{TT}(\textbf{k}) = \frac{1}{2}\int\!\!\int \!d^{D-1}\textbf{x} \, d^{D-1}\textbf{y}\,  J^{\mu\nu}_{TT}(\textbf{x})J^{\alpha\lambda}_{TT}(\textbf{y}) e^{i\textbf{k}\cdot(\textbf{y}-\textbf{x})},
\end{equation}
\begin{equation}\label{delta_LL}
\mkern-12mu \Delta^{\mu\nu\alpha\lambda}_{LL}(\textbf{k}) = \frac{1}{2}\int\!\!\int \!d^{D-1}\textbf{x} \, d^{D-1}\textbf{y}\,  J^{\mu\nu}_{LL}(\textbf{x})J^{\alpha\lambda}_{LL}(\textbf{y}) e^{i\textbf{k}\cdot(\textbf{y}-\textbf{x})},
\end{equation}
\begin{equation}\label{delta_TL_2}
\mkern-12mu \Delta^{\mu\nu\alpha\lambda}_{TL}(\textbf{k}) = \frac{1}{2}\int\!\!\int \!d^{D-1}\textbf{x} \, d^{D-1}\textbf{y}\,  J^{\mu\nu}_{TL}(\textbf{x})J^{\alpha\lambda}_{TL}(\textbf{y}) e^{i\textbf{k}\cdot(\textbf{y}-\textbf{x})},
\end{equation}
\begin{eqnarray}\label{Upsilon}
\mkern-12mu \Upsilon^{\mu\nu\alpha\lambda}(\textbf{k}) = \frac{1}{2}\int\int \!d^{D-1}\textbf{x} \, d^{D-1}\textbf{y}\, e^{\textbf{k}\!\cdot\!(\textbf{y}-\textbf{x})}  \bigg( J^{\mu\nu}_{TT}(\textbf{x})J^{\alpha\lambda}_{LL}(\textbf{y}) + J^{\mu\nu}_{LL}(\textbf{x})J^{\alpha\lambda}_{TT}(\textbf{y}) \bigg) ,
\end{eqnarray}
\begin{eqnarray}\label{Pi}
\Pi^{\mu\nu\alpha\lambda}(\textbf{k}) \!=\! \frac{1}{2}\int\!\!\!\int d^{D-1}\textbf{x} \, d^{D-1}\textbf{y} \,e^{i\textbf{k}\cdot(\textbf{y}-\textbf{x})} 
\bigg\{ J^{\mu\nu}_{TT}(\textbf{x}) J^{\alpha\lambda}_{TL}(\textbf{y}) +J^{\mu\nu}_{TL}(\textbf{x}) J^{\alpha\lambda}_{TT}(\textbf{y}) + J^{\mu\nu}_{TL}(\textbf{x}) J^{\alpha\lambda}_{LL}(\textbf{y}) \!+\!J^{\mu\nu}_{LL}(\textbf{x}) J^{\alpha\lambda}_{TL}(\textbf{y})\! \bigg\}.
\end{eqnarray}

On the other hand, the most general free Lagrangian for a rank-2 symmetric tensor field $H^{\mu\nu}$ can be written as follows 
\begin{eqnarray}\label{lagrangian_tensor}
\mkern-15mu \mathcal{L}_0 \!=\! \frac{1}{2}\! H^{\mu\nu} \!\bigg( A_1(\Box) P^{(2)}_{\mu\nu,\alpha\lambda} \!+\! A_2(\Box) P^{(1)}_{\mu\nu,\alpha\lambda} \!+\! A_3(\Box) P^{(0-s)}_{\mu\nu,\alpha\lambda}\! +
 A_4(\Box) \!P^{(0-w)}_{\mu\nu,\alpha\lambda} \!+\! A_5(\Box)\! P^{(0-sw)}_{\mu\nu,\alpha\lambda}\! +\! A_5(\Box) \!P^{(0-ws)}_{\mu\nu,\alpha\lambda} \!\bigg) H^{\alpha\lambda},\,\,\,
\end{eqnarray}
where $\{P^{(2)}, P^{(1)},\cdots,P^{(0-ws)}\}$ is the set of Barnes-Rivers operators (see Appendix \ref{BarnesRivers}). The Feynman propagator in momentum space is given in turn by 

\begin{eqnarray}
D(k) = \frac{1}{\tilde{A}_1(k)}P^{(2)} + \frac{1}{\tilde{A}_2(k)}P^{(1)} +\frac{1}{K}\left( \tilde{A}_4(k) P^{(0-s)} + \tilde{A}_3(k) P^{(0-w)} 
-\tilde{A}_5(k)P^{(0-sw)} - \tilde{A}_5(k) P^{(0-ws)}\right) ,
\end{eqnarray}
where $K = \tilde{A}_3(k)\tilde{A}_4(k) - \tilde{A}_5(k)^2$. 
 
It follows that
\begin{eqnarray}
D_{\mu\nu\alpha\lambda}(\textbf{k}) \Delta^{\mu\nu\alpha\lambda}_{TT}(\textbf{k}) \!=\! \frac{1}{D\!-\!1}\bigg[ \frac{\tilde{A}_4(\textbf{k})}{K} \eta_{\mu\nu}\eta_{\alpha\lambda}+  \frac{(D\!-\!1) \eta_{\mu\alpha}\eta_{\nu\lambda}- \eta_{\mu\nu}\eta_{\alpha\lambda}}{\tilde{A}_1(\textbf{k})} \bigg]\Delta^{\mu\nu\alpha\lambda}_{TT}(\textbf{k}),
\end{eqnarray}
\begin{eqnarray}
D_{\mu\nu\alpha\lambda}(\textbf{k}) \Delta^{\mu\nu\alpha\lambda}_{TL}(\textbf{k}) = \frac{1}{\tilde{A}_2(\textbf{k})} \eta_{\mu\alpha}\eta_{\nu\lambda} \Delta^{\mu\nu\alpha\lambda}_{TL}(\textbf{k}),
\end{eqnarray}
\begin{eqnarray}
D_{\mu\nu\alpha\lambda}(\textbf{k}) \Delta^{\mu\nu\alpha\lambda}_{LL}(\textbf{k}) = \frac{\tilde{A}_3(\textbf{k})}{K} \eta_{\mu\nu}\eta_{\alpha\lambda} \Delta^{\mu\nu\alpha\lambda}_{LL}(\textbf{k}),
\end{eqnarray}
\begin{eqnarray}
\mkern-12mu D_{\mu\nu\alpha\lambda}(\textbf{k}) \Upsilon^{\mu\nu\alpha\lambda}(\textbf{k}) = \frac{-\tilde{A}_5(\textbf{k})}{\sqrt{(D-1)}K} \eta_{\mu\nu}\eta_{\alpha\lambda} \Upsilon^{\mu\nu\alpha\lambda}(\textbf{k}),
\end{eqnarray}
\begin{eqnarray}
D_{\mu\nu\alpha\lambda}(\textbf{k}) \Pi^{\mu\nu\alpha\lambda}(\textbf{k}) = 0.
\end{eqnarray}
Substitution of these  results into (\ref{potential_tensor}), furnishes the  expression
\begin{eqnarray}
\mkern-25mu \!\!\! & E_D& \!\!\!= \bigg(\eta_{\mu\alpha}\eta_{\nu\lambda}\!\!-\!\! \frac{\eta_{\mu\nu}\eta_{\alpha\lambda}}{D-1} \bigg)\! \int \frac{d^{D-1}\textbf{k}}{(2\pi)^{D-1}}\, \frac{\Delta^{\mu\nu\alpha\lambda}_{TT}(\textbf{k})}{\tilde{A}_1(\textbf{k})} + \eta_{\mu\alpha}\eta_{\nu\lambda}\int \frac{d^{D-1}\textbf{k}}{(2\pi)^{D-1}}\, \frac{\Delta^{\mu\nu\alpha\lambda}_{TL}(\textbf{k})}{\tilde{A}_2(\textbf{k})} +  \!\!\!\!\\ &+& \!\!\!\! \frac{\eta_{\mu\nu}\eta_{\alpha\lambda}}{D-1}\int \frac{d^{D-1}\textbf{k}}{(2\pi)^{D-1}}\, \frac{\tilde{A}_4(\textbf{k})}{K}\Delta^{\mu\nu\alpha\lambda}_{TT}(\textbf{k}) \!+\! \eta_{\mu\nu}\eta_{\alpha\lambda}\int \frac{d^{D-1}\textbf{k}}{(2\pi)^{D-1}}\, \frac{\tilde{A}_3(\textbf{k})}{K}\Delta^{\mu\nu\alpha\lambda}_{LL}(\textbf{k}) \!-\!\frac{\eta_{\mu\nu}\eta_{\alpha\lambda}}{\sqrt{D-1}}\int \frac{d^{D-1}\textbf{k}}{(2\pi)^{D-1}}\, \frac{\tilde{A}_5(\textbf{k})}{K}\Upsilon^{\mu\nu\alpha\lambda}(\textbf{k}). \nn 
\end{eqnarray}
Taking this result into account, we can split the potential energy as follows
\begin{equation} \label{energy_complete}
E_D = E_D^{(2)} + E_D^{(1)} + E_D^{(0-s)} + E_D^{(0-w)} + E_D^{(sw)},
\end{equation}
where  the following definitions  were used
\begin{eqnarray}\label{ENERGY_S_2}
E_D^{(2)} \!=\! \bigg(\eta_{\mu\alpha}\eta_{\nu\lambda}\!\!-\!\! \frac{\eta_{\mu\nu}\eta_{\alpha\lambda}}{D-1} \bigg)\! \int \frac{d^{D-1}\textbf{k}}{(2\pi)^{D-1}}\, \frac{\Delta^{\mu\nu\alpha\lambda}_{TT}(\textbf{k})}{\tilde{A}_1(\textbf{k})},
\end{eqnarray}
\begin{eqnarray}\label{ENERGY_S_1}
E_D^{(1)} \!=\! \eta_{\mu\alpha}\eta_{\nu\lambda}\int \frac{d^{D-1}\textbf{k}}{(2\pi)^{D-1}}\, \frac{\Delta^{\mu\nu\alpha\lambda}_{TL}(\textbf{k})}{\tilde{A}_2(\textbf{k})},
\end{eqnarray}
\begin{eqnarray}\label{ENERGY_S_0-s}
E_D^{(0-s)} \!=\! \frac{\eta_{\mu\nu}\eta_{\alpha\lambda}}{D-1}\int \frac{d^{D-1}\textbf{k}}{(2\pi)^{D-1}}\, \frac{\tilde{A}_4(\textbf{k})}{K}\Delta^{\mu\nu\alpha\lambda}_{TT}(\textbf{k}),
\end{eqnarray}
\begin{eqnarray}\label{ENERGY_S_0-w}
E_D^{(0-w)} \!=\! \eta_{\mu\nu}\eta_{\alpha\lambda}\int \frac{d^{D-1}\textbf{k}}{(2\pi)^{D-1}}\, \frac{\tilde{A}_3(\textbf{k})}{K}\Delta^{\mu\nu\alpha\lambda}_{LL}(\textbf{k}),
\end{eqnarray}
\begin{eqnarray}\label{ENERGY_S_sw}
E_D^{(sw)} \!=\! -\frac{\eta_{\mu\nu}\eta_{\alpha\lambda}}{\sqrt{D-1}}\int \frac{d^{D-1}\textbf{k}}{(2\pi)^{D-1}}\, \frac{\tilde{A}_5(\textbf{k})}{K}\Upsilon^{\mu\nu\alpha\lambda}(\textbf{k}),
\end{eqnarray}

The   energy $E_D$ was split  in several parts so that   the contributions coming from each sector of the propagator $D_{\mu\nu\alpha\lambda}$ were separated.  In our notation, the superscript in $E_D^{(I)}$ ($I = 2,1,0-s,0-w,sw$) identifies this correspondence between the propagator sectors and the energy contribution.

Now, let us particularize this expression for  external currents $J_{TT}^{\mu\nu}$, $J_{TL}^{\mu\nu}$ and $J_{LL}^{\mu\nu}$ related to  two point-like charges. 
\begin{eqnarray}\label{current_TT}
\mkern-12mu J^{\mu\nu}_{TT}(\textbf{x}) \!=\! \eta^{\mu 0}\!\eta^{\nu 0}\!\bigg(\! Q^{(TT)}_1 \delta^{d}(\textbf{x}\! -\! \textbf{a}_1)\! +\!  Q^{(TT)}_2 \delta^{d}(\textbf{x}\! -\! \textbf{a}_2)\! \bigg),\,
\end{eqnarray}
\begin{eqnarray}\label{current_TL}
\!\!\!\!\! J^{\mu\nu}_{TL}(\textbf{x}) \!\!=\!\! \frac{1}{\sqrt{2}}\eta^{\mu 0}\!\pt^\nu\!\bigg(\! Q^{(TL)}_1 \delta^{d}(\textbf{x}\! -\! \textbf{a}_1)\! +\!  Q^{(TL)}_2 \delta^{d}(\textbf{x}\! -\! \textbf{a}_2)\! \bigg) + \frac{1}{\sqrt{2}}\eta^{\nu 0}\!\pt^\mu\!\bigg(\! Q^{(TL)}_1 \delta^{d}(\textbf{x}\! -\! \textbf{a}_1)\! +\!  Q^{(TL)}_2 \delta^{d}(\textbf{x}\! -\! \textbf{a}_2)\! \bigg),\,
\end{eqnarray}
\begin{eqnarray}\label{current_LL}
\mkern-12mu J^{\mu\nu}_{LL}(\textbf{x}) \!=\! \pt^{\mu}\!\pt^{\nu}\!\bigg(\! Q^{(LL)}_1 \delta^{d}(\textbf{x}\! -\! \textbf{a}_1)\! +\!  Q^{(LL)}_2 \delta^{d}(\textbf{x}\! -\! \textbf{a}_2)\! \bigg),\,
\end{eqnarray}
where $d=D-1$ is the number of spatial dimensions. 

Substitution of Eqs. (\ref{current_TT}), (\ref{current_TL}) and (\ref{current_LL}) into the set of equations (\ref{delta_TT})-(\ref{Upsilon}), allows to write 

\begin{eqnarray}
\Delta^{\mu\nu\alpha\lambda}_{TT}(\textbf{k}) = \eta^{\mu 0}\eta^{\nu 0}\eta^{\alpha 0}\eta^{\lambda 0} Q_1^{(TT)}Q_2^{(TT)} \cos(\textbf{k}\cdot\textbf{r}),
\end{eqnarray}
\begin{eqnarray}
\mkern-20mu\Delta^{\mu\nu\alpha\lambda}_{TL}(\textbf{k}) \!=\! \bigg(\eta^{\nu 0}\eta^{\alpha 0} \delta^\mu_i \delta^\lambda_j \!+\! \eta^{\mu 0}\eta^{\alpha 0} \delta^\nu_i \delta^\lambda_j \!+\! \eta^{\mu 0}\eta^{\lambda 0} \delta^\nu_i \delta^\alpha_j +\,\eta^{\nu 0}\eta^{\lambda 0} \delta^\mu_i \delta^\alpha_j \bigg) \frac{Q_1^{(TL)}Q_2^{(TL)}}{2} k^i k^j \cos(\textbf{k}\cdot\textbf{r}),\,\,
\end{eqnarray}
\begin{eqnarray}
\mkern-5mu \Delta^{\mu\nu\alpha\lambda}_{LL}(\textbf{k}) \!=\! \delta^{\mu}_i\delta^{\nu}_j\delta^{\alpha}_m \delta^{\lambda}_l k^i k^j k^m k^l \, Q_1^{(LL)}\!Q_2^{(LL)}\! \cos(\textbf{k}\!\cdot\!\textbf{r}),
\end{eqnarray}
\begin{eqnarray}
\mkern-10mu \Upsilon^{\mu\nu\alpha\lambda}(\textbf{k}) = -\eta^{\mu 0}\eta^{\nu 0} \delta^\alpha_i \delta^\lambda_j k^i k^j \, \bigg(Q_1^{(TT)}Q_2^{(LL)} + Q_1^{(LL)} Q_2^{(TT)}\bigg) \cos(\textbf{k}\cdot\textbf{r}) ,
\end{eqnarray}
where $\textbf{r}=a_2-a_1$. 

Using the above equations in the set of Eqs. (\ref{ENERGY_S_2})-(\ref{ENERGY_S_sw}), we come to the conclusion  that
\begin{eqnarray}\label{ENERGY_S_2_final}
E_D^{(2)} \!=\! \frac{D-2}{D-1} Q_1^{(TT)}Q_2^{(TT)} \int \frac{d^{D-1}\textbf{k}}{(2\pi)^{D-1}}\, \frac{e^{i\textbf{k}\cdot\textbf{r}}}{\tilde{A}_1(\textbf{k})},
\end{eqnarray}
\begin{eqnarray}\label{ENERGY_S_1_final}
E_D^{(1)} \!=\! -Q_1^{(TL)}Q_2^{(TL)} \int \frac{d^{D-1}\textbf{k}}{(2\pi)^{D-1}}\, \frac{\textbf{k}^2 e^{i \textbf{k}\cdot \textbf{r}}}{\tilde{A}_2(\textbf{k})},
\end{eqnarray}
\begin{eqnarray}\label{ENERGY_S_0-s_final}
E_D^{(0-s)} \!=\! \frac{Q_1^{(TT)}Q_2^{(TT)}}{D-1}\int \frac{d^{D-1}\textbf{k}}{(2\pi)^{D-1}}\, \frac{\tilde{A}_4(\textbf{k})}{K}e^{i\textbf{k}\cdot\textbf{r}},
\end{eqnarray}
\begin{eqnarray}\label{ENERGY_S_0-w_final}
E_D^{(0-w)} \!=\! Q_1^{(LL)}Q_2^{(LL)}\int \frac{d^{D-1}\textbf{k}}{(2\pi)^{D-1}}\, \frac{\tilde{A}_3(\textbf{k})}{K}(\textbf{k}^2)^2e^{i\textbf{k}\cdot\textbf{r}},
\end{eqnarray}
\begin{eqnarray}\label{ENERGY_S_sw_final}
E_D^{(sw)} \!=\! -\frac{Q_1^{(TT)}Q_2^{(LL)}}{\sqrt{D-1}}\int \frac{d^{D-1}\textbf{k}}{(2\pi)^{D-1}}\, \frac{\tilde{A}_5(\textbf{k})}{K} \textbf{k}^2 e^{i\textbf{k}\cdot\textbf{r}} - \frac{Q_1^{(LL)}Q_2^{(TT)}}{\sqrt{D-1}}\int \frac{d^{D-1}\textbf{k}}{(2\pi)^{D-1}}\, \frac{\tilde{A}_5(\textbf{k})}{K} \textbf{k}^2 e^{i\textbf{k}\cdot\textbf{r}},
\end{eqnarray}

\section{Prescription for computing the non-relativistic interparticle potential energy mediated by  antisymmetric rank-2 tensor fields \label{antissimetrico}}

Consider now an antisymmetric tensor field ($A_{\mu\nu} = -A_{\nu\mu}$). For convenience's sake, 
we restrict our discussion  to $D=4$. In the case of interactions mediated by antisymmetric tensors, the external current   is an antisymmetric rank-2 tensor, which we denote by $j^{\mu\nu}$. Since this antisymmetric rank-2 tensor carries two different kinds of spin-1 representations, labeled here by $[1b]$ and $[1e]$, we may   divide  the external current into two parts $j^{\mu\nu} = j^{\mu\nu}_{e} + j^{\mu\nu}_{b}$. It is remarkable that we can write $j^{\mu\nu}_{e} = \pt^\mu j_e^{\nu}-\pt^\nu j_e^{\mu}$  and $j^{\mu\nu}_{b} = \varepsilon^{\mu\nu\alpha\lambda}\pt_\alpha j^b_\lambda$, where $j_{e}^\mu$ represents a conserved vector current and $j^b_\lambda$ stands for a pseudo-vectorial current.

 The potential energy is given by
\begin{eqnarray}\label{potential_tensor_antisym}
\mkern-5muE_{D=4} \!=\! \int \frac{d^{3}\textbf{k}}{(2\pi)^{3}}\, D_{\mu\nu\alpha\lambda}(\textbf{k}) \bigg( \Delta^{\mu\nu\alpha\lambda}_{e}(\textbf{k}) \!+\! \Delta^{\mu\nu\alpha\lambda}_{b}(\textbf{k}) \bigg)\!, \,\,
\end{eqnarray}
where the following  judicious definitions were utilized
\begin{equation}\label{delta_e}
\mkern-12mu \Delta^{\mu\nu\alpha\lambda}_{e}(\textbf{k}) = \frac{1}{2}\int\!\!\int \!d^{3}\textbf{x} \, d^{3}\textbf{y}\,  j^{\mu\nu}_{e}(\textbf{x})j^{\alpha\lambda}_{e}(\textbf{y}) e^{i\textbf{k}\cdot(\textbf{y}-\textbf{x})},
\end{equation}

\begin{equation}\label{delta_b}
\mkern-12mu \Delta^{\mu\nu\alpha\lambda}_{b}(\textbf{k}) = \frac{1}{2}\int\!\!\int \!d^{3}\textbf{x} \, d^{3}\textbf{y}\,  j^{\mu\nu}_{b}(\textbf{x})j^{\alpha\lambda}_{b}(\textbf{y}) e^{i\textbf{k}\cdot(\textbf{y}-\textbf{x})}.
\end{equation}

On the other hand, the most  general free Lagrangian  for an antisymmetric rank-2 tensor field 
can be cast in the form
\begin{equation} \label{lagrangian_anti}
\mathcal{L}_0 = \frac{1}{2}A^{\mu\nu} \left( B_e(\Box) P^{[1e]}_{\mu\nu,\alpha\lambda} + B_b(\Box) P^{[1b]}_{\mu\nu,\alpha\lambda} \right)A^{\alpha\lambda} ,
\end{equation}
where $\{P^{[1b]},P^{[1e]}\}$ is the set of antisymmetric Barnes-Rivers spin projectors (see Appendix \ref{BarnesRivers}).
 
The Feynman propagator in  momentum space is given in turn by  
\begin{equation}
D_{\mu\nu\alpha\lambda} = \frac{1}{\tilde{B}_e(k)} P^{[1e]}_{\mu\nu,\alpha\lambda} + \frac{1}{\tilde{B}_b(k)} P^{[1b]}_{\mu\nu,\alpha\lambda}.
\end{equation}
Substitution of  the last expression into (\ref{delta_e}) and (\ref{delta_b}), furnishes the result 

\begin{eqnarray}
D_{\mu\nu\alpha\lambda}(\textbf{k}) \Delta^{\mu\nu\alpha\lambda}_{e}(\textbf{k}) = \frac{1}{\tilde{B}_e(\textbf{k})} \eta_{\mu\alpha}\eta_{\nu\lambda} \Delta^{\mu\nu\alpha\lambda}_{e}(\textbf{k}),
\end{eqnarray}
\begin{eqnarray}
D_{\mu\nu\alpha\lambda}(\textbf{k}) \Delta^{\mu\nu\alpha\lambda}_{b}(\textbf{k}) = \frac{1}{\tilde{B}_b(\textbf{k})} \eta_{\mu\alpha}\eta_{\nu\lambda} \Delta^{\mu\nu\alpha\lambda}_{b}(\textbf{k}),
\end{eqnarray}
 Dividing  now  the potential energy into two contributions, we find
\begin{eqnarray}
E_{D=4} = E_{D=4}^{[1e]} + E_{D=4}^{[1b]},
\end{eqnarray}
where 
\begin{eqnarray}\label{ENERGY_1e}
E_{D=4}^{[1e]} = \eta_{\mu\alpha}\eta_{\nu\lambda} \int \frac{d^{3} \textbf{k}}{(2\pi)^{3}} \frac{1}{\tilde{B}_e(\textbf{k})} \Delta^{\mu\nu\alpha\lambda}_{e}(\textbf{k}),
\end{eqnarray}

\begin{eqnarray}\label{ENERGY_1b}
E_{D=4}^{[1b]} = \eta_{\mu\alpha}\eta_{\nu\lambda} \int \frac{d^{3} \textbf{k}}{(2\pi)^{3}} \frac{1}{\tilde{B}_b(\textbf{k})} \Delta^{\mu\nu\alpha\lambda}_{b}(\textbf{k}).
\end{eqnarray}
Following Ref. \cite{Barone}, we consider    external currents  concentrated along two parallel $d$-dimensional branes ($d\leq2$). The case $d=0$ leads to the usual point-like distributions. We chose   the following   external currents 
\begin{eqnarray}\label{current_e}
\mkern-25mu   j^{\mu\nu}_e \!=\! \frac{\pt^\mu \!\eta^{\nu 0}}{\sqrt{2}}\bigg(\!Q^{e}_1 \delta^{(3\!-\!d)}(\textbf{x}_{\perp} \!-\! \textbf{a}_1)\! +\! Q^{e}_2 \delta^{(3\!-\!d)}(\textbf{x}_{\perp} \!-\! \textbf{a}_2)\bigg) \!-\!\!\frac{\pt^\nu \!\eta^{\mu 0}}{\sqrt{2}}\bigg(\!Q^{e}_1 \delta^{(3\!-\!d)}(\textbf{x}_{\perp} \!-\! \textbf{a}_1) \!+ \!Q^{e}_2 \delta^{(3\!-\!d)}(\textbf{x}_{\perp} \!-\! \textbf{a}_2)\!\bigg),
\end{eqnarray}

\begin{eqnarray} \label{current_b}
\mkern-5mu j^{\mu\nu}_b \!\!=\!\! \frac{\varepsilon^{\mu\nu\alpha\lambda} \pt_\lambda}{\sqrt{2}}  \bigg(\! V_\alpha\delta^{(3\!-\!d)}(\textbf{x}_{\perp}\! \!-\!\! \textbf{a}_1) \!+\! W_\alpha\delta^{(3\!-\!d)}(\textbf{x}_{\perp} \!-\! \textbf{a}_2)\!\bigg)\!,\,
\end{eqnarray}
where $\textbf{x}_\perp$ stands for spatial coordinates orthogonal to the branes, $\textbf{a}_1$ and $\textbf{a}_2$ represent the position vector along the branes, $Q^{e}$ is a charge associated with the representation $[1e]$ and $V_\alpha$ and $W_\alpha$ are pseudo four vectors that are  constant in the reference frame where the calculations are made. Using Eqs. (\ref{current_e}) and (\ref{current_b}), respectively, in (\ref{ENERGY_1e}) and (\ref{ENERGY_1b}), we obtain the following expressions for the potential energy mediated by an antisymmetric rank-2 tensor associated with two parallel $d$-branes
\begin{eqnarray}\label{ENERGY_1e_final}
E_{D=4}^{[1e]} = -\,Q^{e}_1 \,Q^{e}_2 \int \frac{d^{n}\textbf{k}_\perp}{(2\pi)^{n}}  \frac{\textbf{k}_\perp^2 \,e^{i\textbf{k}_\perp \cdot \textbf{r}_\perp}}{\tilde{B}_e(\textbf{k}_\perp)} ,
\end{eqnarray}
and
\begin{eqnarray}\label{ENERGY_1b_final}
\mkern-5mu E_{D=4}^{[1b]} = \int\!\! \frac{d^{n}\textbf{k}_\perp}{(2\pi)^{n}} \frac{(\textbf{k}_\perp \!\cdot\! \textbf{V})(\textbf{k}_\perp \!\cdot\! \textbf{W}) \!+\! \textbf{k}_\perp^2 W^\mu V_\mu}{\tilde{B}_b(\textbf{k}_\perp)}e^{i\textbf{k}_\perp \cdot \textbf{r}_\perp} ,
\end{eqnarray}
where $n=3-d$ and $\textbf{r}_\perp = (\textbf{a}_2 - \textbf{a}_1)_\perp$ is the projection of the vector $\textbf{r}$ along the orthogonal direction to the branes.

\section{Computing the interparticle potential energy for some specific models \label{examples}}
 To text the simplicity and efficacy of the prescription we have built out, we discuss below specific examples concerning the computation of the  nonrelativistic interparticle  
 potential energy related to  the cases developed in Secs. II-IV.
\subsection{Spin-0 particle described by a vector field  \label{EX1}}

In the usual approach, spin-$0$ particles are described by the usual Klein-Gordon scalar field, however other descriptions are possible, for instance, vector field representations \cite{PRD_renato_1}. The motivation to consider vectorial representation of spin-$0$ particles is twofold: (i) Vector field possess a more intricate structure than the scalar one, hence, would be expected the possibilities of new vertex (when interacting with other fields) which are not contemplated by the usual scalar description; (ii) One can use the vectorial description for spin-$0$ particles as a theoretical laboratory to explore a systematic procedure to construct more elaborated dual theories.
 
In this section we consider a model that  describes a massive spin-0 particle via a vectorial representation, being its  dynamics  governed by the  Lagrangian \cite{PRD_renato_1}
\begin{equation} \label{spin-0_2nd_1}
\mathcal{L}^{(0,m)}_A = \frac{1}{2} \left[ (\pt_\mu A^\mu)^2 - m^2 A_\mu A^\mu \right],
\end{equation} 

\noindent which can be rewritten as 
\begin{equation}
\mathcal{L}^{(0,m)}_A = \frac{1}{2} A^\mu \left[ - m^2 \theta_{\mu\nu} - (\Box + m^2)\omega_{\mu\nu} \right] A^\nu .
\end{equation}

Therefore, $\tilde{a}(\textbf{k}) = -m^2$ and $\tilde{b}(\textbf{k}) = (k^2 - m^2)|_{k^0=0} = - (\textbf{k}^{2} + m^2)$. A quick glance   at the corresponding Feynman propagator (see Eq. (\ref{propagator_vector})) clearly shows that $A^\mu$ only propagates in the longitudinal sector;  as a consequence, we can get rid of the potential energy contribution coming from  the transverse external current. Accordingly, the only contribution to the $D$-dimensional potential energy between point-like charges  comes from the longitudinal current (\textit{i.e.}, $E_D = E_D^L$). Consequently, the $D$-dimensional potential energy can be computed via the expression

\begin{equation}
E_D =  \frac{Q_1^{(L)}Q_2^{(L)}}{(2\pi)^{D-1}}\,\int\,d^{D-1}\textbf{k} \,\frac{\textbf{k}^2 \,e^{i\textbf{k}\cdot\textbf{r}}}{\textbf{k}^2 + m^2} .
\end{equation}

Performing the integration we obtain
\begin{equation} \label{Energy_exp_1}
E_D(r) = -\frac{Q_1^{(L)}Q_2^{(L)}}{(2\pi)^{(D-1)/2}} \bigg( \frac{m^{D+1}}{r^{D-3}} \bigg)^{1/2} K_{\frac{D-3}{2}}(mr).
\end{equation}
\noindent where $K_\nu$ is the modified Bessel function of  the second  order of the order $\nu$.

For $D=4$, (74) reduces to

\begin{equation} \label{potential_4D}
E_{D=4}(r) = -\frac{Q_1^{(L)}Q_2^{(L)}}{4\pi} \frac{m^2 e^{-mr}}{r}.
\end{equation}

\subsection{Spin-0 particle described by a higher-derivative vector field  \label{EX2}}

We move now to a system in which  a spin-0 particle  is described by  a higher-derivative vector field. In this case we have, essentially, the same motivation of the previous section. The model  is defined by the Lagrangian (see \cite{PRD_renato_1})
\begin{equation}\label{spin-0_4nd_1}
\mathcal{L}^{(0,m)}_B = -\frac{1}{2} \left[ \pt_\mu B^\mu (\Box + m^2)\pt_\nu B^\nu \right] .
\end{equation} 
However, in the present model we have a peculiarity that we cannot forget to mention. The presence of a $\Box^2$ term in the above Lagrangian may lead to the misconception of assume the presence of a ghost state. However, it was demonstrated in Ref. \cite{PRD_renato_1} that the above Lagrangian is ghost-free. In fact, there are other examples in the literature in which the presence of higher derivatives do not imply the presence of ghost states, for instance, the so-called ``new massive gravity" provide a ghost-free description for higher derivative gravity in $D = 2+1$ \cite{NMG}.

Note that the above Lagrangian is gauge invariant under the transformation $B^\mu \to B^{\mu} + \pt_\nu \Lambda^{\mu\nu}$, with $\Lambda^{\mu\nu} = -\Lambda^{\nu\mu}$. It is astonishing  that this gauge symmetry implies that $J^{\mu}_T = \theta^\mu_\nu J^\nu = 0$;  as a result, the interparticle potential energy has only contributions coming from the longitudinal part. After the addition of a gauge fixing term (We  work with the gauge fixing term $-\lambda F_{\mu\nu}(B)^2$, where $F_{\mu\nu}(B) = \pt_\mu B_\nu - \pt_\nu B_\mu$.),  the Lagrangian (\ref{spin-0_4nd_1}) assumes the form 
\begin{equation}
\mathcal{L}^{(0,m)}_B = \frac{1}{2} B^\mu \left[ \Box(\Box + m^2)\omega_{\mu\nu}  + \lambda \Box\theta_{\mu\nu} \right] B^\nu ,
\end{equation}
where $\lambda$ is a gauge fixing parameter. Consequently, $\tilde{a}(\textbf{k}) = -\lambda k^2|_{k^0 = 0} = \lambda \textbf{k}^2$ and $\tilde{b}(\textbf{k}) = (k^2 - m^2)k^2 |_{k^0 = 0} = (\textbf{k}^2 + m^2)\textbf{k}^2$. Thence, using Eq. {\ref{potential_vector_L}}, we get
\begin{equation}
E_D =  -\frac{Q_1^{L}Q_2^{L}}{(2\pi)^{D-1}}\, \,\int\,d^{D-1}\textbf{k}\,\frac{e^{i\textbf{k}\cdot\textbf{r}}}{\textbf{k}^2 + m^2}  .
\end{equation}
Performing the integral, we obtain a result identical to that found  in the last section, namely
\begin{equation} \label{Energy_exp_2}
E_D(r) = -\frac{Q_1^{L}Q_2^{L}}{(2\pi)^{\frac{D-1}{2}}}\bigg( \frac{m}{r} \bigg)^{{(D-3)}/2} K_{\frac{D-3}{2}}(mr).
\end{equation}
It is interesting to observe that in this example the vector field must have mass dimension $(D-4)/2$ (in order to obtain an action with mass dimension zero) and, as a consequence, the charge $Q^{(L)}$ has mass dimension $(4-D)/2$. So, despite  the apparent difference between Eqs. (\ref{Energy_exp_1}) and (\ref{Energy_exp_2}), both expressions give the the correct dimension for the potential energy, \textit{i.e.},  $+1$.\\
\indent For $D=4$, (79) leads to the result
\begin{equation}
E_{D=4} = -\frac{Q_1^{(L)}Q_2^{(L)}}{4\pi} \frac{e^{-mr}}{r}.
\end{equation}
It is worth noticing that  in this case we have a well defined energy  as  $m\to0$. 

\subsection{Spin-1 particle described by a symmetric rank-2 tensor field \label{EX3}}

In the same sense that spin-$0$ particles may be described scalar or vector fields, one can use a tensorial representation for spin-$1$ instead of the usual vector field. The same kind of motivations mentioned in Sec. \ref{EX1} may be applied to the present situation. In fact, since a symmetric rank-2 tensor has a more intricate structure than the vectorial representation, we could expect to new vertex possibilities. 

We analyze a model in which a symmetric rank-2 tensor field describes a massive spin-1 particle. The dynamics of the system  is governed by the  Lagrangian \cite{PRD_renato_1,PRD_renato_2}
\begin{eqnarray}
\mathcal{L}^{(1,m)}_W = -(\pt^\nu W_{\mu\nu})^2 + \frac{m^2}{2} \left( W_{\mu\nu}^2 - \frac{W^2}{D-1} \right),
\end{eqnarray}
where $W_{\mu\nu}$ is a symmetric tensor with mass dimension $(D-2)/2$. After some  algebraic manipulations, the preceding  Lagrangian  can be rewritten as
\begin{eqnarray}
\mathcal{L}^{(1,m)}_W \!\!=\!\!\frac{1}{2} W^{\mu\nu}\bigg[ m^2 P^{(2)}_{\mu\nu,\alpha\lambda} -(m^2 + \Box) P^{(1)}_{\mu\nu,\alpha\lambda} + \left( \frac{D-2}{D-1}m^2 + 2 \Box \right) P^{(0-w)}_{\mu\nu,\alpha\lambda}  \!\!-\!\! \frac{m^2}{\sqrt{D-1}} \bigg( P^{(0-sw)}_{\mu\nu,\alpha\lambda}+ P^{(0-ws)}_{\mu\nu,\alpha\lambda} \bigg) \bigg]W^{\alpha\lambda}.
\end{eqnarray}
Comparing this expression with (\ref{lagrangian_tensor}), the following   correspondences can be made
\begin{eqnarray}
A_1(\Box) = m^2 \, \Rightarrow \, \tilde{A}_1(\textbf{k}) = m^2,
\end{eqnarray}
\begin{eqnarray}
A_2(\Box) = m^2 +\Box \, \Rightarrow \, \tilde{A}_2(\textbf{k}) = m^2 + \textbf{k}^2,
\end{eqnarray}
\begin{eqnarray}
A_3(\Box) = 0 \, \Rightarrow \, \tilde{A}_3(\textbf{k}) = 0 ,
\end{eqnarray}
\begin{eqnarray}
\mkern-10mu A_4(\Box) = \frac{D\!-\! 2}{D\!-\! 1}m^2 \!+\! 2 \Box \, \Rightarrow \, \tilde{A}_4(\textbf{k}) = \frac{D\!-\! 2}{D\!-\! 1}m^2 \!+\! 2 \textbf{k}^2,
\end{eqnarray}
\begin{eqnarray}
A_5(\Box) = - \frac{m^2}{\sqrt{D-1}} \, \Rightarrow \, \tilde{A}_5(\textbf{k}) = - \frac{m^2}{\sqrt{D-1}}. 
\end{eqnarray}
It is easy to show that the only propagating mode describes a massive spin-1 particle (see Refs. \cite{PRD_renato_1,PRD_renato_2}); consequently, the unique  contribution relevant to the interparticle potential energy is $E_D^{(1)}$, given by Eq. (\ref{ENERGY_S_1_final}). Substitution of $\tilde{A}_2(\textbf{k}) = m^2 + \textbf{k}^2$ into (\ref{ENERGY_S_1_final}),  allows to write

\begin{eqnarray}
E_D \!=\! -Q_1^{(TL)}Q_2^{(TL)} \int \frac{d^{D-1}\textbf{k}}{(2\pi)^{D-1}}\, \frac{\textbf{k}^2 e^{i \textbf{k}\cdot \textbf{r}}}{m^2 + \textbf{k}^2}.
\end{eqnarray}
Solving this integral, we obtain
\begin{equation} \label{Energy_exp_3}
E_D(r) = \frac{Q_1^{(TL)}Q_2^{(TL)}}{(2\pi)^{(D-1)/2}} \bigg( \frac{m^{D+1}}{r^{D-3}} \bigg)^{1/2} K_{\frac{D-3}{2}}(mr).
\end{equation}
Note that the charge $Q^{(TL)}$ has mass dimension $(2-D)/2$ which implies that the potential has the correct mass dimension $+1$. \\
\indent In particular, the case $D=4$ leads to the  result
\begin{equation} 
E_{D=4}(r) = \frac{ Q_1^{(TL)}Q_2^{(TL)}}{4\pi} \frac{m^2 e^{-mr}}{r}.
\end{equation}

\subsection{Spin-1 particle described by a symmetric rank-2 tensor field  containing higher-derivatives \label{EX4}}

We discuss now  a second example of our prescription for  symmetric rank-2 tensor fields:  a higher derivative model which describes a spin-1 particle via a tensor representation. Once again the presence of higher derivatives may lead to the wrong conclusion that the present model contains ghost states. In Ref. \cite{PRD_renato_1} it was shown that this model is absent of ghosts, hence this is an interesting example of unitary higher derivative theories describing massive spin-$1$ particles. 

The model under consideration is defined by the following Lagrangian \cite{PRD_renato_1,Tese_Renato}
\begin{eqnarray}
\mathcal{L}^{(1,m)}_H = -\frac{1}{4}F_{\mu\nu}^2[\pt H] + \frac{m^2}{2} (\pt_\nu H_{\mu\nu})^2,
\end{eqnarray}
where $F_{\mu\nu}[\pt H] = \pt_\mu (\pt^\alpha H_{\nu\alpha}) - \pt_\nu (\pt^\alpha H_{\mu\alpha})$ and $H_{\mu\nu}$ is a symmetric tensor field with mass dimension $(D-4)/2$. According with reference \cite{PRD_renato_1}, after the introduction of a gauge fixing term (We  work with the gauge fixing term $\lambda G_{\mu\nu}^2$, where $G_{\mu\nu}(H) = \Box H_{\mu\nu} - \pt^\alpha \pt_{\mu} H_{\nu\alpha} - \pt^\alpha \pt_{\nu} H_{\mu\alpha} + \eta_{\mu\nu} \pt^\alpha \pt^\beta H_{\alpha \beta}$ and $\lambda$ is a gauge parameter.), this  Lagrangian may be recast as below
\begin{eqnarray}
\mathcal{L}^{(1,m)}_H = \frac{1}{2} H^{\mu\nu} \bigg( 2\lambda \Box P^{(2)}_{\mu\nu\alpha\lambda} - \Box(\Box+m^2)P^{(1)}_{\mu\nu\alpha\lambda}  + 2 \lambda \Box^4 P^{(0-s)}_{\mu\nu\alpha\lambda} - 2 m^2 \Box P^{(0-w)}_{\mu\nu\alpha\lambda}\bigg) H^{\alpha\lambda}.
\end{eqnarray}
Comparing the last equation with $\ref{lagrangian_tensor}$, we obtain
\begin{eqnarray}
A_1(\Box) = \lambda \Box^4 \Rightarrow \tilde{A}_1(\textbf{k}) = 2 \lambda (\textbf{k}^2)^4,
\end{eqnarray}
\begin{eqnarray}
A_2(\Box) = -\Box(\Box + m^2) \Rightarrow \tilde{A}_2(\textbf{k}) = -\textbf{k}^2(\textbf{k}^2+m^2),
\end{eqnarray}
\begin{eqnarray}
A_3(\Box) = 2 \lambda \Box^4 \Rightarrow \tilde{A}_3(\textbf{k}) = 2 \lambda (\textbf{k}^2)^4,
\end{eqnarray}
\begin{eqnarray}
A_4(\Box) = 2 m^2 \Box \Rightarrow \tilde{A}_4(\textbf{k}) = 2m^2 \textbf{k}^2,
\end{eqnarray}
and $A_5(\Box) = 0$. In ref. \cite{PRD_renato_1} is argued that the gauge symmetry presented in this model ensures that the only relevant contribution for the external current is $J^{\mu\nu}_{TL}$; consequently the unique contribution for the potential energy is $E_D^{(1)}$. Using Eq. (\ref{ENERGY_S_1_final}), we find
\begin{eqnarray}
E_D \!=\! Q_1^{(TL)}Q_2^{(TL)} \int \frac{d^{D-1}\textbf{k}}{(2\pi)^{D-1}}\, \frac{e^{i \textbf{k}\cdot \textbf{r}}}{m^2 + \textbf{k}^2}.
\end{eqnarray}
Solving the this integral, we arrive at the conclusion that
\begin{equation} \label{Energy_exp_4}
E_D(r) = \frac{Q_1^{(TL)}Q_2^{(TL)}}{(2\pi)^{(D-1)/2}} \bigg( \frac{m}{r} \bigg)^{(D-3)/2} K_{\frac{D-3}{2}}(mr).
\end{equation}
Some remarks regarding the mass dimension of the charge $Q^{(TL)}$ are in order here: (i)  The charge $Q^{(TL)}$ has mass dimension $(4-D)/2$. (ii)  Despite  the apparent difference between Eqs. (\ref{Energy_exp_3}) and (\ref{Energy_exp_4}), both equations have the correct mass dimension as far as the the nonrelativistic potential energy is concerned.\\ 
\indent In particular, for  $D=4$ (98) assumes the form 
\begin{equation} 
E_{D=4}(r) = \frac{ Q_1^{(TL)}Q_2^{(TL)}}{4\pi} \frac{e^{-mr}}{r}.
\end{equation}

\subsection{Spin-0 particle described by Kalb-Ramond \label{EX5}}

In order to exemplify our method developed to compute the interparticle potential associated with interactions mediated by antisymmetric rank-2 tensor we first consider the Kalb-Ramond (KR) field \cite{KR}. The KR field is an alternative description for spin-$0$ objects. Since the KR field is described by an antisymmetric rank-2 tensor it allows a large number of interactions than the usual scalar field representation  for spin-$0$ particles. For instance, the KR field can mediated interactions between extended objects, which is very relevant in the context of string theories.

Let us start from the KR Lagrangian
\begin{equation}
\L_{K\!R} = \frac{1}{6} G_{\mu\nu\alpha}G^{\mu\nu\alpha},
\end{equation}
where $G_{\mu\nu\alpha} = \pt_\mu B_{\nu\alpha} + \pt_\nu B_{\alpha\mu} + \pt_\alpha B_{\mu\nu}$ and $B_{\mu\nu}(= -B_{\nu\mu})$ is the antisymmetric KR field. The KR model is invariant under the following gauge transformation $B_{\mu\nu} \rightarrow B_{\mu\nu} + \pt_\mu \Lambda_\nu - \pt_\nu \Lambda_\nu$ and, as a consequence, this symmetry leads to the conservation equation $\pt_\mu j^{\mu\nu} = 0$, where $j^{\mu\nu}$ is an antisymmetric external current. In order to apply our prescription to compute the interparticle potential energy let us rewrite the KR Lagrangian (argumented with the gauge fixing term $\L_{g.f.} = \frac{1}{2\beta} (\pt_\mu B^{\mu\nu})^2 $) in the usual field-operator-field form
\begin{equation}
\L_{K\!R} = \frac{1}{2}B^{\mu\nu} \left[ -\Box P^{[1b]}_{\mu\nu\alpha\beta} - \frac{1}{2 \beta} \Box \,P^{[1e]}_{\mu\nu\alpha\beta}  \right] B^{\alpha \beta}.
\end{equation}
Comparing the last expression with (\ref{lagrangian_anti}), we promptly find
\begin{eqnarray}
& B_e(\Box) =  - \frac{1}{2 \beta} \Box \Rightarrow \tilde{B}_e(\textbf{k}) =  - \frac{1}{2 \beta} \textbf{k}^2,&\\
& B_b(\Box) = - \Box \Rightarrow \tilde{B}_b(\textbf{k}) = - \textbf{k}^2 .&
\end{eqnarray}

Remarkably, the conservation equation $\pt_\mu j^{\mu\nu} = 0$, which is a consequence of the gauge invariance of the KR model, exclude the possibility of the contribution $j_e^{\mu\nu}$ in the external source and, as a consequence, the only relevant part for the interparticle potential is the $E_D^{[1b]}$ contribution. Accordingly, using Eq. (\ref{ENERGY_1b_final}) along with above expression for $\tilde{B}_b(\textbf{k})$, we are lead to the following result
\begin{eqnarray}
E_{D=4}^{[1b]} =  - \int \frac{d^{n}\textbf{k}_\perp}{(2\pi)^{n}} \frac{(\textbf{k}_\perp \!\cdot\! \textbf{V})(\textbf{k}_\perp \!\cdot\! \textbf{W})}{\textbf{k}_\perp^2}e^{i\textbf{k}_\perp \cdot \textbf{r}_\perp}  - \int \frac{d^{n}\textbf{k}_\perp}{(2\pi)^{n}} \frac{\textbf{k}_\perp^2 W^\mu V_\mu}{\textbf{k}_\perp^2}e^{i\textbf{k}_\perp \cdot \textbf{r}_\perp}.
\end{eqnarray}
Evaluating the above integral we are lead to the following result
\begin{equation}
E_{D=4}^{[1b]} = -\frac{2^{(n-2)/2}}{(2\pi)^{n/2}} \Gamma(n/2) \left( \frac{\textbf{V}\cdot\textbf{W}}{r_\perp^n } - n \frac{ (\textbf{V}\cdot\textbf{r}_\perp ) (\textbf{W}\cdot\textbf{r}_\perp ) }{r_\perp^{n+2}} \right),
\end{equation} 
where we have discarded the contribution of the second integral, since it would be relevant only for $\textbf{r} = 0$. The result obtained for the interparticle potential associated with de KR field using our prescription is in complete agreement with the result obtained in Ref. \cite{Barone}, where the interparticle potential were computed by means of more standard methods.

\subsection{Spin-1 particle described by an antisymmetric rank-2 tensor \label{EX6}}

As the last example we study a model in which  a massive spin-1 particle is described by an antisymmetric rank-2 tensor. The corresponding Lagrangian is given by
\begin{eqnarray} \label{lagrangian_antissimetrico}
\mathcal{L}^{(m,1)}_B = -(\pt_\mu B^{\mu\nu})^2 + \frac{m^2}{2}B^{\mu\nu}B_{\mu\nu}.
\end{eqnarray}
This system was analyzed in references \cite{Ecker,Kampf} within the context of strong interaction in order to provide an effective description of the low-energy regime of QCD. 

Here we are interested in finding its nonrelativistic potential energy. After some algebraic manipulations, we may recast the above Lagrangian as follows
\begin{eqnarray}
\mathcal{L}^{(m,1)}_B = \frac{1}{2}B^{\mu\nu}\bigg[ (\Box + m^2)P^{[1e]}_{\mu\nu\alpha\lambda} \!+\! m^2P^{[1b]}_{\mu\nu\alpha\lambda} \bigg]B^{\alpha\lambda}.
\end{eqnarray}
Comparing the last equation with (\ref{lagrangian_anti}), we promptly find
\begin{eqnarray}
& B_e(\Box) = \Box + m^2 \Rightarrow \tilde{B}_e(\textbf{k}) = \textbf{k}^2 + m^2,&\\
& B_b(\Box) = m^2 \Rightarrow \tilde{B}_b(\textbf{k}) = m^2.&
\end{eqnarray}
It is straightforward  to conclude that $[1e]$ is the only sector which has  a particle content; thence, the relevant contribution for the potential energy is $E_D^{[1e]}$. Accordingly,
\begin{eqnarray}
E_{D=4}^{[1e]} = -\,Q^{e}_1 \,Q^{e}_2 \int \frac{d^{3-d}\textbf{k}_\perp}{(2\pi)^{3-d}}  \frac{\textbf{k}_\perp^2 \,e^{i\textbf{k}_\perp \cdot \textbf{r}_\perp}}{\textbf{k}_\perp^2 + m^2} ,
\end{eqnarray}
Solving this  integral, we obtain
\begin{eqnarray}
E_{D=4}^{[1e]} = \frac{\,Q^{e}_1 \,Q^{e}_2}{(2\pi)^{(3-d)/2}} \bigg( \frac{m^{5-d}}{r_\perp^{1-d}} \bigg)^{1/2} K_{\frac{1-d}{2}}(m r_\perp).
\end{eqnarray}
For a point-like charge, \textit{i.e.}, a  $0$-brane, we obtain  a Yukawa-like potential
\begin{eqnarray}
E_{D=4} = \frac{ Q^{e}_1 \,Q^{e}_2}{4\pi }\frac{m^2 e^{-m r}}{r},
\end{eqnarray}  
which  clearly exhibits  a repulsive behavior for like  charges, as expected. 

\section{Concluding Remarks \label{conclusion}}
 Based on a procedure recently built out by Accioly {\it et al.} \cite{Accioly1}  --- which allows a straightforward computation of the $D$-dimensional nonrelativistic interparticle potential energy for electromagnetic models --- and relaxing the condition assumed by them concerning the conservation of  the external current, a general  extension of their method was found which, among other things, contemplates dual models.

The main points we have analyzed are listed below.
 \begin{enumerate}
\item  As far as the interactions mediated by vector fields are concerned, the possibility of propagating modes were considered both  in longitudinal (spin-0) and transverse (spin-1) sectors.

\item For interactions mediated by  a rank-2 symmetric tensor field, we took into account that a symmetric tensor can be split into spin-2, spin-1 and two spin-0 sectors which, as a result, allowed different couplings between the tensor field and external currents. 
 \item We  extended the method we have constructed so that    interactions mediated by rank-2 antisymmetric tensors could be included. Although we have restricted  our  computations to $D=4$,  a very simple algorithm for  computing the potential energy was obtained. 
 \end{enumerate}
Last but not least, we would like to comment on the limit $m\to0$ concerning the examples analyzed in this work. Before going on, we remark that interesting discussions regarding the massless limit of spin-1 models can be found in Refs. \cite{PRD_renato_2,Tese_Renato}. Coming back to our main theme, we call attention to the fact that in sections \ref{EX1}, \ref{EX3} and \ref{EX6} there is a global multiplicative factor $m^2$ in the  potential energy expressions which, at first sight, signs an apparent inconsistency  in the limit $m^2\to0$.  Nevertheless, there exists no inconsistency in these systems. Indeed, it is easy to prove that in the first and second models there is no propagating mode if $m^2=0$; consequently, we may not expect any potential energy in this case. The last example is more sophisticated.  Using the master action technique \cite{Master_action} (see Appendix \ref{MasterAction}),  it is easy to show that if $m^2=0$ the model considered in \ref{EX6} behaves like a spin-0 particle, leading to a finite potential but with a different behavior from that found  in the massless limit of the alluded section. Therefore,  a kind of DVZ discontinuity \cite{vDV,Zakharov} occurs in this limit.

\section*{Acknowledgments}
The authors are grateful to CNPq for financial support. GPB thanks  professors A. Accioly and J. Helay\"{e}l-Neto for  reading  the manuscript and  presenting helpful suggestions. GPB is also grateful to C. Marques and P.I.C. Caneda for  fruitful  discussions regarding the Barnes-Rivers operators.  

\appendix
\section{Barnes-Rivers operators \label{BarnesRivers}}

To begin with, we define the transverse and longitudinal vectorial  projector operators as follows
\begin{eqnarray}
\theta_{\mu\nu} = \eta_{\mu\nu} - \frac{\pt_\mu \pt_\nu}{\Box} \quad \textmd{and} \quad \omega_{\mu\nu} = \frac{\pt_\mu \pt_\nu}{\Box},
\end{eqnarray}
which satisfy the trivial algebra
\begin{eqnarray}
\theta^2 = \theta , \quad \omega^2 = \omega \quad \textmd{and} \quad \theta \omega = \omega \theta = 0. 
\end{eqnarray}
 
Using these operators, we may write the complete set of the $D$-dimensional Barnes-Rivers operators \cite{Accioly1,Rivers} as
\begin{equation}
P^{(2)}_{\mu\nu,\alpha\lambda} = \frac{1}{2}(\theta_{\mu\alpha} \theta_{\nu\lambda} + \theta_{\mu\lambda} \theta_{\nu\alpha}) - \frac{1}{D-1} \theta_{\mu\nu} \theta_{\alpha\lambda} ,
\end{equation}
\begin{equation}
P^{(1)}_{\mu\nu,\alpha\lambda} = \frac{1}{2}(\theta_{\mu\alpha} \omega_{\nu\lambda} + \theta_{\mu\lambda} \omega_{\nu\alpha} + \theta_{\nu\alpha} \omega_{\mu\lambda} + \theta_{\nu\lambda} \omega_{\mu\alpha}),
\end{equation}
\begin{equation}
P^{(0-s)}_{\mu\nu,\alpha\lambda} = \frac{1}{D-1} \theta_{\mu\nu} \theta_{\alpha\lambda},
\end{equation}
\begin{equation}
P^{(0-w)}_{\mu\nu,\alpha\lambda} = \omega_{\mu\nu} \omega_{\alpha\lambda},
\end{equation}
\begin{equation}
P^{(0-sw)}_{\mu\nu,\alpha\lambda} = \frac{1}{\sqrt{D-1}}\theta_{\mu\nu} \omega_{\alpha\lambda},
\end{equation}
\begin{equation}
P^{(0-ws)}_{\mu\nu,\alpha\lambda} = \frac{1}{\sqrt{D-1}}\omega_{\mu\nu} \theta_{\alpha\lambda},
\end{equation}
which satisfy a very useful algebra having the following  non-vanishing products 
\begin{eqnarray}
&(P^{(2)})^2 = P^{(2)}, \quad (P^{(1)})^2 = P^{(1)},\quad (P^{(0-s)})^2 = P^{(0-s)},&\nn \\
&(P^{(0-w)})^2= P^{(0-w)} ,\quad P^{(0-s)}P^{(0-sw)} = P^{(0-sw)}, &\nn \\
& P^{(0-w)} P^{(0-ws)} = P^{(0-ws)}, \quad P^{(0-sw)}P^{(0-w)} = P^{(0-sw)},&\nn \\
&P^{(0-ws)}P^{(0-ws)} = P^{(0-s)}, \quad P^{(0-ws)} P^{(0-s)} = P^{(0-ws)} ,&\nn\\
&P^{(0-ws)}P^{(0-sw)} = P^{(0-w)}.&
\end{eqnarray}
On the other hand,  the set of  antisymmetric four-dimensional Barnes-Rivers operator is given by \cite{Helayel} 
\begin{equation}
P^{[1b]}_{\mu\nu,\alpha\lambda} = \frac{1}{2}(\theta_{\mu\alpha} \theta_{\nu\lambda} - \theta_{\mu\lambda} \theta_{\nu\alpha}),
\end{equation}
\begin{equation}
P^{[1e]}_{\mu\nu,\alpha\lambda} = \frac{1}{2}(\theta_{\mu\alpha} \omega_{\nu\lambda} + \theta_{\nu\lambda} \omega_{\mu\alpha} - \theta_{\mu\lambda} \omega_{\nu\alpha} - \theta_{\nu\alpha} \omega_{\mu\lambda}),
\end{equation}
which satisfy the very simple algebra
\begin{eqnarray}
&(P^{[1b]})^2 = P^{[1b]}, \quad (P^{[1e]})^2 = P^{[1e]},&\nn\\
&P^{[1b]}P^{[1e]} = P^{[1e]}P^{[1b]}=0.
\end{eqnarray}

\section{Master action and the physical content of the example \ref{EX6} \label{MasterAction}}
  To demonstrate the equivalence between the massive antisymmetric spin-1 model found in section \ref{EX6} and the Maxwell-Proca theory, we appeal to the master action technique \cite{Master_action}. In this  vein, the master action related to  those models can be written as
\begin{eqnarray}
S_m &=& \int d^Dx \bigg( m^2 B_{\mu \nu}B^{\mu \nu} +  2m(\pt^\mu B_{\mu \nu})A^{\nu}  +\frac{1}{2}m^2A_{\nu}A^\nu \bigg).
\end{eqnarray}
 Performing  a Gaussian integration in (B1) with respect to the vector field $A_{\mu}$,   we obtain an action proportional to (\ref{lagrangian_antissimetrico}).

 On the other hand, we can  also integrate (B1) with respect to the antisymmetric tensor field $B_{\mu \nu}$.  However, before doing this, we integrate by parts the aforementioned action, which leads to the result 

\begin{eqnarray}
\!\!S_m\!=\!\!\int \!d^Dx [m^2B_{\mu \nu}B^{\mu \nu}-mB_{\mu \nu}F^{\mu \nu}+\frac{1}{2}m^2A_\nu A^\nu ].
\end{eqnarray}
\noindent The $F^{\mu \nu}$ tensor field that appears in (B2)  is the familiar Maxwell electromagnetic tensor $F^{\mu \nu}=\pt^\mu A^\nu-\pt^\nu A^{\mu}$. 

It is easy to see  that an integration with respect to $B_{\mu \nu}$  of (B2) reproduces the Maxwell-Proca action. \\
\indent A natural question can then be posed at this point: What is  is the physical content of the massless limit of the massive spin-1 field dealt with in section \ref{EX6}? To answer  this question we use the  master action
\begin{eqnarray}
S_0=\int d^Dx\{C_\nu C^\nu-2C^\nu(\pt^\mu B_{\mu \nu})\},
\end{eqnarray}
\noindent where the vector $C_\nu$ is an auxiliary vector field. 

Integrating  this result with respect to $C_\nu$, we arrive at a model proportional to the massless limit of  that given by Eq. (\ref{lagrangian_antissimetrico}). Before integrating over $B_{\mu \nu}$, we  perform an integration by parts to obtain the  expression
\begin{eqnarray} \label{Action_C}
S_0=\int d^Dx\{C_\nu C^\nu+F^{\mu \nu}_CB_{\mu \nu}\},
\end{eqnarray}

\noindent where we have defined $F^{\mu \nu}_C=\pt^\mu C^{\nu}-\pt^\nu C^{\mu}$. Now, an integration with respect to $B_{\mu \nu}$ can be understood as a functional Dirac's delta which generates the constraint $F^{\mu \nu}_C = 0$. The  general solution of this equation is given by $C_{\nu} = \partial_\nu \phi$, where $\phi$ is a scalar field. Using this solution in (\ref{Action_C}), we arrive at the following result
\begin{eqnarray}
S_0=\int d^Dx\{\pt_\mu \phi \pt^\mu \phi \}.
\end{eqnarray}
\indent A quick inspection of  $S_m$ and  $S_0$ allows us to conclude that there is a discontinuity in the degrees of freedom  of the massless limit of the antisymmetric rank-2 tensor field used in Section \ref{EX6}. The massive case describes a spin-1 particle and  the massless one   a spin-0 field. This may be related to the ``spin jumping" phenomena discussed in \cite{spin_jump}.


\begin{thebibliography}{99}
\footnotesize
\bibitem{Accioly1} A. Accioly, J. Helay\"{e}l-Neto, F.E. Barone, F.A. Barone and P. Gaete, Phys. Rev. D \textbf{90}, 105029 (2014).

\bibitem{Accioly2} A. Accioly, J. Helay\"{e}l-Neto, F.E. Barone and W. Herdy, Class. Quantum Grav. \textbf{32}, 035021 (2015).

\bibitem{Duality} S. E. Hjelmeland and U. Lindstrom, \textit{Duality for the Non-Specialist
}, arXiv:hep-th/9705122.

\bibitem{Zee} A. Zee, \textit{Quantum Field Theory in a Nutshell}, 2nd ed. (Princeton University Press, Princeton, NJ, 2010).

\bibitem{Barone} F.A. Barone, F.E. Barone and J. Helay\"{e}l-Neto, Phys. Rev. D \textbf{84}, 065026 (2011).

\bibitem{PRD_renato_1} D. Dalmazi and R.C. Santos, Phys. Rev. D \textbf{87}, 085021 (2013).

\bibitem{NMG} E. Bergshoeff, O. Hohm, and P. K. Townsend, Phys. Rev. Lett. \textbf{102}, 201301 (2009).

\bibitem{PRD_renato_2} D. Dalmazi and R.C. Santos, Phys. Rev. D \textbf{84}, 045027 (2011).

\bibitem{Tese_Renato} R.C. Santos, Part\'{i}culas de spin-1 em D-dimens\~{o}es via tensor sim\'{e}trico, MSc dissertation, 2012, S\~{a}o Paulo State University (In Portuguese).


\bibitem{KR} M. Kalb and P. Ramond, Phys. Rev. D \textbf{9}, 2273 (1973).

\bibitem{Ecker} G. Ecker, J. Gasser, H. Leutwyler, A. Pich and E. De Rafael, Phys. Lett. B \textbf{223}, 425 (1989).

\bibitem{Kampf} K. Kampf, J. Novotný, J. Trnka, Acta Physica Polonica B \textbf{38}, 9 (2007).

\bibitem{Master_action} S. Deser and R. Jackiw, Phys. Lett. B \textbf{139}, 371 (1984).

\bibitem{vDV} H. van Dam and M. Veltman, Nucl. Phys. B \textbf{22}, 397 (1970).

\bibitem{Zakharov} V.I. Zakharov, JETP Lett. \textbf{12}, 312 (1970).

\bibitem{Rivers} R.J. Rivers Nuovo Ciment \textbf{34}, 387 (1964).

\bibitem{Helayel} F.A. Gomes Ferreira, P.C. Malta, L.P.R. Ospedal and J.A. Helay\"{e}l-Neto, Eur. Journ. Phys. C \textbf{75}, 238 (2015).

\bibitem{spin_jump} A. Smailagic and E. Spallucci, Journ. of Phys. A: Math. Gen. \textbf{34}, L435 (2001). 
\end{thebibliography}
\end{document}